\documentclass[twocolumn,prl,superscriptaddress,showpacs,floatfix]{revtex4}
\usepackage{times}
\usepackage{xspace}
\usepackage{graphicx}
\usepackage{color}
\usepackage{amsmath, amsthm, amssymb}

\begin{document}
\setlength\arraycolsep{2pt}

%\title{Mesoscopic simulations of membranes and membrane associated proteins: application to
%the mobility of integral membrane proteins with intrinsic curvature}

\title{Hybrid elastic/discrete-particle approach to biomembrane dynamics with application 
to the mobility of curved integral membrane proteins}

\author{Ali Naji}% \email{E-mail: anaji@chem.ucsb.edu}}
					
\affiliation{Department of Chemistry and Biochemistry \& Department of Physics}
%University of California, Santa Barbara, CA 93106, USA}

%\affiliation{Department of Chemistry and Biochemistry, \& Materials Research Laboratory,
%University of California, Santa Barbara, CA 93106, USA}

\author{ Paul J. Atzberger}

\affiliation{Department of Mathematics, University of California, Santa Barbara, CA 93106, USA}
                                                                
\author{Frank L. H. Brown}% \email{E-mail: flbrown@chem.ucsb.edu}}
\affiliation{Department of Chemistry and Biochemistry \& Department of Physics}
%University of California, Santa Barbara, CA 93106, USA}

\begin{abstract}
We introduce a simulation strategy to consistently couple continuum biomembrane dynamics 
to the motion of discrete biological macromolecules residing within or on the membrane.  The
methodology is used to study the diffusion of integral membrane proteins that impart
a curvature on the bilayer surrounding them.  Such proteins exhibit a substantial 
reduction in diffusion coefficient relative to ``flat" proteins; this effect is explained by
elementary hydrodynamic considerations.
 \end{abstract}


\maketitle

%%%%%%%%%%%%%%%%%%%%%%%%%%%%%%%%%%%%%%%%%%%%%%%%%%%%%%%%

Lipid-bilayer membranes are among the most important and most versatile 
components of biological cells \cite{textbook,gennis}; biomembranes 
protect cells from their surroundings, provide a means to compartmentalize 
subcellular structures (and the functions of these structures) 
and act as a scaffolding for countless biochemical reactions involving 
membrane associated proteins.  Our conceptual picture of biological
membranes as a ``two-dimensional oriented solution of integral proteins ... in the
viscous phospholipid bilayer"\cite{singer} was popularized well over thirty years ago.  Quantitative
physical models for the energetics \cite{helfrich} and dynamics \cite{Brochard} associated
with shape fluctuations of homogeneous fluid membranes and for the lateral diffusion coefficient of 
integral membrane proteins within a flat bilayer \cite{Saffmann} were developed shortly
thereafter and still find widespread use up to this day.

\begin{figure}[t!]
\begin{center}
\includegraphics[angle=0,width=7.cm]{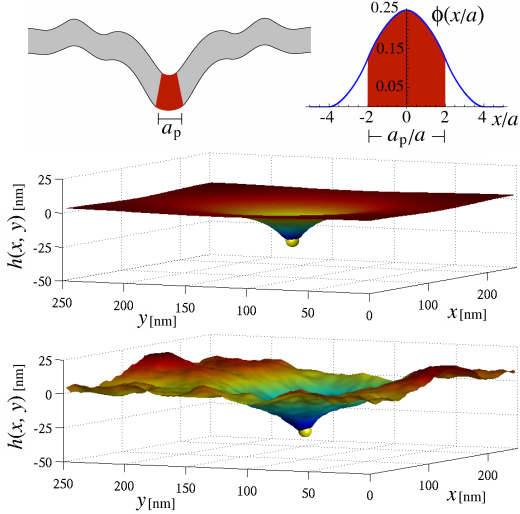}
\caption{
(Color online) An appropriately shaped protein will tend to distort the local shape of the bilayer
into a bent configuration.  %Within 
In our simulations, the protein's effective size is defined
via the envelope function  
%defined in  Eq. (\ref{eq:delta}) and displayed graphically next to the cartoon (top).  
 in  Eq. (\ref{eq:delta}) (displayed next to the cartoon, top). 
The distortion is most readily seen by minimizing
$\mathcal{H}$ for the composite protein-bilayer system (middle), however the
protein's influence is strong enough to maintain visibly apparent perturbations, even
in the presence of %thermal 
fluctuations (bottom).  See Table \ref{tab:sim_para} for parameters. }
%The physical parameters correspond 
%to those specified in the text
%with $C_{\mathrm{p}} =0.1\,{\mathrm{nm}}^{-1}$, $K_{\mathrm{p}}/K_{\mathrm{m}}=8$.}
\vspace{-.7cm}
\label{fig:snapshot}
\end{center}
\end{figure}

Interestingly, the coupling of protein diffusion to the shape of
the membrane surface has become a subject of study only relatively
recently (see \cite{Halle_NMR,Seifert_pre,Naji} and references within).  
One well studied consequence of membrane shape fluctuations is
that a protein must travel a longer distance between two points in 3D space 
if the paths connecting these points are constrained to lie on a rough surface as opposed
to a flat plane.  This purely geometric effect is expected to have practical
experimental implications; measurements that capture a 2D projection of
the true motion over the membrane surface will infer diffusion coefficients of diminished
magnitude relative to the intrinsic lateral diffusion locally tangent to the bilayer surface
\cite{Halle_NMR,Seifert,Gov_pre,Seifert_pre,Naji}.  Beyond this generic effect, which should apply
to anything moving on the membrane surface in relatively passive fashion (lipids, proteins, choleseterol, etc.), certain
membrane associated proteins effect shape changes in the bilayer.  Specific examples
include the SERC1a calcium pump \cite{serc1a} and BAR (Bin, Amphiphysin, Rvs) domain
dimers \cite{bar_science}.  Direct structural evidence from X-ray crystallography \cite{serc1a,bar_science}, experimental studies of vesicle topology at the micron scale \cite{Girard,bar_science}, and atomically detailed simulations \cite{voth} all indicate the ability of these proteins to drive membrane curvature (see Fig. \ref{fig:snapshot}).  

The diffusion of membrane proteins with intrinsic curvature is more complex than the 
diffusion of a relatively passive spectator and remains incompletely explored in the literature.
Although stochastic differential equations coupling the lateral motion of curved proteins
to thermal shape fluctuations of a continuous elastic bilayer have been proposed \cite{Seifert,Seifert_crv}, these equations have only been analyzed under the simplifying 
assumption that the protein does not affect the shape of the membrane surface.  Under such 
an approximation, it was predicted both analytically \cite{Seifert} and numerically \cite{Seifert_crv} that
curved proteins are expected to diffuse more rapidly than flat ones.  The full numerical 
analysis provided in the present work suggests exactly the opposite effect--protein curvature 
decreases lateral mobility across the bilayer.  The disagreement with previous work is
attributable to the fact that the protein's influence on bilayer shape is  of primary importance
(see Fig. \ref{fig:snapshot}) and cannot be ignored.

%This Letter describes a consistent approach to  
%couple continuum level membrane dynamics with the motion of discrete molecules
%along the membrane surface.   The proposed methodology is capable of treating general
%interactions between a membrane surface and molecular/supermolecular structures
%within the cell and should find wide use in varied biophysical studies.  Hybrid continuum/discrete particle descriptions are natural for many membrane biophysics problems.  This Letter introduces 
%a practical means to study the behavior of such models.

Our starting point is the Monge-gauge Helfrich Hamiltonian \cite{helfrich} 
for energetics of a homogeneous membrane surface under conditions of
vanishing tension
\begin{equation}
{\mathcal H}_0 = \frac{1}{2}\int_{{\mathcal A}_\bot}\! {\mathrm{d}}{\mathbf x}\, 
   \big[K_{\mathrm{m}}\,(\nabla^2 h)^2+2K'_{\mathrm{m}}\,{\mathcal G}(h)\big],
 \label{eq:H0}
\end{equation}
where $h({\mathbf x})\equiv h(x,y)$ describes the local membrane displacement from a flat
reference plane at $z=0$ (see Fig. \ref{fig:snapshot}).  Here, ${\mathcal G}(h) = (\partial_{xx}h\partial_{yy}h-\partial_{xy}h\partial_{xy}h)$ is the Gaussian curvature and 
$K_{\mathrm{m}}$ and $K'_{\mathrm{m}}$ are the membrane bending modulus and
saddle-splay modulus, respectively.  The integration region, ${\mathcal A}_\bot=L^2$,
is always taken to be a square box with periodic boundary conditions assumed.  We 
treat membrane proteins as localized regions of enhanced rigidity within the bilayer.
A single protein centered within the bilayer at position 
$({\mathbf r}(t),h({\mathbf r} (t))) \equiv (x(t),y(t),h(x(t),y(t)))$ is thus assumed to modify the
Hamiltonian as ${\mathcal H} = {\mathcal H}_0 + {\mathcal H}_{\mathrm{int}}$
with
 \begin{eqnarray}
 {\mathcal H}_{\mathrm{int}} &=& \frac{1}{2}\int_{{\mathcal A}_\bot}\!\!\! 
 	{\mathrm{d}}{\mathbf x}\, G_{\mathrm{p}}({\mathbf x}-{\mathbf r}(t))\,
  					\big[K_{\mathrm{p}}\,(\nabla^2 h-2C_{\mathrm{p}})^2 
					  \label{eq:V_crv} \\
	&&\qquad \qquad \qquad- K_m\,(\nabla^2 h)^2 + 2(K'_{\mathrm{p}}-K'_{\mathrm{m}})\,{\mathcal G}(h) \big]. \nonumber
\end{eqnarray}
$K_{\mathrm{p}}$ and $K'_{\mathrm{p}}$ are the protein bending and saddle-splay moduli and
$C_{\mathrm{p}}$ is the %preferred (``spontaneous") 
spontaneous curvature associated with protein shape.
The protein shape function $G_{\mathrm{p}}({\mathbf x}-{\mathbf r}(t))$ describes the 
envelope of protein influence over bilayer elastic properties; the specific function
chosen will be discussed in detail below.  
%(Association of the protein with the bilayer is
%implicit within this description, since the $x,y$ coordinates of the protein impose the
%z coordinate through the condition $z=h(x,y)$.)
%such that $\int_{{\mathcal A}_\bot}\! {\mathrm{d}}{\mathbf x}\, G_{\mathrm{p}}({\mathbf x}-{\mathbf r}(t)) = A_{\mathrm{p}}$. 

Using a Fourier representation $h({\mathbf x}) = 
\frac{1}{L^2}\sum_{{\mathbf q}} h_{{\mathbf q}}\, e^{{\mathrm{i}} {\mathbf q}\cdot{\mathbf x}}$, 
membrane dynamics may be cast as a set of coupled Langevin equations for the 
individual Fourier modes \cite{Granek,Brown_prl}
\begin{equation}
\dot{h}_{{\mathbf q}}(t)  =  \Lambda_{\mathbf{q}}F_{{\mathbf q}}(t)
 				+ \sqrt{ k_{\mathrm{B}}TL^2\Lambda_{{\mathbf q}}}\, \,\xi_{{\mathbf q}}(t),
 		\label{eq:q_memb_langevin}
\end{equation}
where $F_{{\mathbf q}}(t)$ is the Fourier-transform of the force per unit area
 $F({\mathbf x}, t)  = -\delta {\mathcal H}/\delta h({\mathbf x}, t)$, $\Lambda_{{\mathbf q}}=1/(4\eta \,q)$
corresponds to the  Oseen hydrodynamic kernel $\Lambda({\mathbf x}) = 1/(8\pi \eta |{\mathbf x}|)$, 
and $\xi_{{\mathbf q}}(t)$ is a Gaussian white noise with 
$\langle \xi_{{\mathbf q}}(t)\rangle =0$ and $\langle \xi_{{\mathbf q}}(t) \, \xi_{{\mathbf q'}}(t') \rangle = 
2\delta_{{\mathbf q}, -{\mathbf q}'} \,\delta(t-t')$, to ensure satisfaction of the fluctuation-dissipation
theorem. 

%Eq. \ref{eq:q_memb_langevin} should be regarded as a practical means to implement
%Brownian dynamics with hydrodynamic interactions \cite{mccammon} for membrane shape fluctuations.

%The force may be decomposed into the free elastic part and the interaction contribution as 
%$F_{{\mathbf q}} = - \Omega_{\mathbf{q}}h_{\mathbf{q}} + F_{{\mathbf q}}^{\mathrm{int}}$,
%where $\Omega_{\mathbf{q}} = K q^4$ is the energy spectrum of a free membrane and $F_{{\mathbf q}}^{\mathrm{int}}$
%may be evaluated numerically  for a given protein-membrane configuration and interaction  potential $ {\mathcal H}_{\mathrm{int}}$. 

%%%------

The protein's position may similarly be described in terms of a Langevin
equation  that implicitly enforces protein localization to the membrane surface \cite{Naji,Seifert_pre}.
The two independent variables describing this motion are the components
of ${\mathbf r}(t)$ (with $i,j=1,2$ and summation convention assumed)
 \begin{equation}
   \dot{r}_i(t) = D_0 \,v_i + \sqrt{2D_0}\,\tau_{ij}\, \eta_j(t) + \frac{D_0}{k_{\mathrm{B}} T}(g^{-1})_{ij} f_j. 
   \label{eq:xy_Langevin_int}
\end{equation}
Here $(g^{-1})_{ij} = \delta_{ij} - \partial_i h \,\partial_j h/g$ is the inverse metric tensor, 
$g=  1 + (\nabla h)^2$ is the determinant of the metric $ g_{ij} = \delta_{ij} + \partial_i h \,\partial_j h$  and
we have defined $\partial_i\equiv \partial/\partial x_i$. 
%The latter determines the area rescaling of surface element upon projection, {\em i.e.} 
%${\mathrm{d}}\mathcal{A}=\sqrt{g}\,{\mathrm{d}}\mathcal{A}_{\bot}$. 
Equation (\ref{eq:xy_Langevin_int}) introduces geometric factors including
$v_i = - \big[(g^{-1})_{jk} \, \partial_j \partial_k h \big]\,
 \partial_i h/{g}$ and $\tau_{ij} = \delta_{ij} - \partial_i h\, \partial_j h/(g + \sqrt{g})$ 
 (i.e., the square-root of the 
  inverse metric tensor $(g^{-1})_{ij} = \tau_{ik} \,\tau_{jk}$). 
%  On the local
% tangent plane, a non-interacting particle ($f_i =0$) exhibits a free Brownian motion 
% with a curvilinear diffusion coefficient $D_0$
% driven by 
The Gaussian white-noise $\eta_i(t)$ 
 with $\langle \eta_i(t)\rangle =0$ and $\langle \eta_i(t)\, \eta_j(t')\rangle = \delta_{ij}\,\delta(t-t')$ 
 guarantees that a tracer particle (defined by $\mathcal{H}_{\mathrm{int}}=0$) undergoes
 a curvilinear random walk over the membrane surface with diffusion coefficient $D_0$.
The last term in Eq. (\ref{eq:xy_Langevin_int}) reflects the effect of the interaction-induced force
$f_i = - \partial {\mathcal H}_{\mathrm{int}}/\partial x_i$.

Equations (\ref{eq:H0})-(\ref{eq:xy_Langevin_int}) specify the stochastic thermal evolution
for a single curved protein coupled to an elastic membrane.  As noted above, equations
very similar to these have been proposed previously \cite{Seifert,Seifert_crv}, but
have only been analyzed by neglecting the contribution of $\mathcal{H}_{\mathrm{int}}$
to $F_{\mathbf q}$ while maintaining its contribution to $f_i$.  To avoid this 
uncontrolled approximation, it is necessary
to introduce a numerical algorithm that can consistently couple
protein position ${\mathbf r}(t)$ to membrane undulations $h({\mathbf x},t)$.  

%%%%---Table of parameters
\begin{table*}[t]
\begin{center}
\begin{tabular}{ c | c | c | c | c | c | c | c | c | c | c | c}
    \hline 
  parameter & box  & lattice  & & protein & temperature & bare protein & bilayer  &
  protein  & saddle-splay  & protein  & solvent \\
   & dimension & spacing & & area & & diffusion & bending & bending & moduli & spontaneous & viscosity \\
   &                      &               &  &          & & coefficient& modulus&modulus&              & curvature \\
  \hline \hline
 symbol &   $L$ & $a$ & $M=\frac{L}{a}$ & $A_{\mathrm{p}}$ &  $T$ 
    		& $D_0$  & $K_{\mathrm{m}}$ & $K_{\mathrm{p}}$  & $K'_{\mathrm{m(p)}}$ & $C_{\mathrm{p}}$ & $\eta$   \\ 
		\hline\hline
  value & 250nm & 2.5nm & 100 & 100${\mathrm{nm}}^2$ &  300${\mathrm K}$ & 5.0$\frac{\mu{\mathrm m}^2}{{\mathrm s}}$ 
    	 & 5$k_{\mathrm{B}}T$ & 40$k_{\mathrm{B}}T$ & $-K_{\mathrm{m(p)}}$ &  0.1${\mathrm{nm}}^{-1}$ & $\eta_{\mathrm{w}}$=0.001${\mathrm{Pa}}\cdot{\mathrm{s}}$ \\ \hline
%    \hline
\end{tabular}
\caption{Default parameter values used in the simulations. The protein area is chosen as $A_{\mathrm{p}}=(4a)^2$
(giving a protein diameter of $a_{\mathrm{p}}\sim10$nm), the $D_0$ value qualitatively reflects the motion of band 3 protein 
dimers on the surface of human red blood cells %(RBC) 
\cite{Brown_prl}, and $\eta_{\mathrm{w}}$ stands for the viscosity of water.
For simplicity we assume that both bilayer and protein
saddle-splay moduli are equal in magnitude but opposite in sign to the corresponding bending moduli.}
\label{tab:sim_para}
\vspace{-.7cm}
\end{center}
\end{table*}

For both physical and numerical purposes, we must truncate the membrane modes
at some short-distance scale $a$, which can for example be taken to represent the typical 
molecular (lipid) size or bilayer thickness.  We thus
limit the Fourier modes appearing in Eq. (\ref{eq:q_memb_langevin}) 
to ${\mathbf q} = (q_x, q_y) = (2\pi n, 2\pi m)/L$ where $M=L/a$ 
with integer $n, m$ in the range $-M/2<n, m\leq M/2$. In principle, 
this reduced set of modes describes a fully continuous membrane height profile via $h({\mathbf x}) = 
\frac{1}{L^2}\sum_{{\mathbf q}} h_{{\mathbf q}}\, e^{{\mathrm{i}} {\mathbf q}\cdot{\mathbf x}}$
at {\em any} given point ${\mathbf x}$. However, it is computationally advantageous to explicitly
track the membrane height only over the discrete $M\times M$ real-space lattice (defined at positions ${\mathbf x}_\alpha=(p, q)a$
with integer $0\leq p, q< M$) conjugate to the chosen $ h_{{\mathbf q}}$'s via  
Fast-Fourier Transformation \cite{FFTW}.  

The interaction between a fully continuous variable describing protein 
position ${\mathbf r}(t)$ and a discrete representation of the membrane 
height field $\{h({\mathbf x}_{\alpha},t)\}$ poses certain challenges.  A minor issue is that
Eq. (\ref{eq:xy_Langevin_int}) requires the shape of the membrane surface
over the entire $x,y$ plane and not just at the lattice sites $\{{\mathbf x}_{\alpha}\}$.
This problem is readily handled via linear interpolation to obtain $h({\mathbf x},t)$
and the required derivatives at arbitrary ${\mathbf x}$ from the corresponding neighboring 
lattice values \cite{Seifert_pre}.  A more complex problem involves the dynamics of the $h_{\mathbf q}(t)$'s
in Eq. (\ref{eq:q_memb_langevin}).  The forces in this equation include contributions
due to the coupling between protein and bilayer from Eq. (\ref{eq:V_crv}).  The envelope
function $G_{\mathrm{p}}({\mathbf x})$ reflects protein size and is quite localized in real space; the
natural way to deal with Eq. (\ref{eq:V_crv}) (and the related force expressions) 
is to approximate the integral by simple quadrature, i.e. $\int {\mathrm d}{\mathbf x}\, G_{\mathrm{p}}({\mathbf x}-{\mathbf r}(t) ) \mathcal{F}({\mathbf x}) \approx a^2\sum_{\alpha} G_{\mathrm{p}}({\mathbf x}_{\alpha}-{\mathbf r}(t) )\mathcal{F}({\mathbf x}_{\alpha})$ for arbitrary function $\mathcal{F}$.
To define a numerical scheme that is both efficient and accurate, the specific functional form 
chosen for $G_{\mathrm{p}}$ is critical.  Naive continuous choices like 2D Gaussians \cite{Brown_prl,Seifert_crv} lead to an effective normalization (as computed by quadrature) 
that varies with the offset between the envelope center and the discrete lattice.  Piecewise linear
forms for $G_{\mathrm{p}}$ can be defined that suffer no such normalization issue, 
but such functions lead to discontinuous derivatives as the protein-lattice offset changes.  
Both scenarios are unacceptable as these numerical issues lead to a breaking of the homogeneity of
the membrane surface; the protein will tend to favor (or disfavor) lattice sites over other regions
of space.

Our numerical description of coupled membrane-protein dynamics shares features
with the Immersed Boundary formulation of hydrodynamics \cite{peskin}.
In that work, a series of envelope functions are introduced that are continuous, localized
in space and strictly preserve normalization as evaluated by quadrature.  For our purposes,
we take $G_{\mathrm{p}}({\mathbf x}) = (A_{\mathrm{p}}/a^2) \phi(x/a)\,\phi(y/a)$ 
with \cite{peskin}
\begin{equation}
 	\phi\big(2u\big) =\frac{1}{16}\left\{
	\begin{array}{lc}
	             5+2u-\sqrt{-7-12u-4u^2}
              		& { -2\leq u\leq -1,}\\	            
		    3+2u+\sqrt{1-4u-4u^2}
	              & { -1\leq u\leq 0, }\\
		     3-2u+\sqrt{1+4u-4u^2}
		      & { 0\leq u\leq 1, }\\	
		      5-2u-\sqrt{-7+12u-4u^2}
		      	& { 1\leq u\leq 2, }
      \end{array}
        \right.  \label{eq:delta}
\end{equation}
and zero for all other $u$ values  (see Fig. \ref{fig:snapshot} for a plot).  $A_{\mathrm p}$ defines an effective protein area and the envelope
function is non-vanishing over a total of 64 lattice sites.  
%Choices involving a significantly
%smaller support region are possible (e.g. 16 lattice sites); Eq. (\ref{eq:delta})
%insures fully converged results for all of the cases considered in this work.
In order to simulate the dynamics of the system, we evolve Eqs. (\ref{eq:q_memb_langevin}) and (\ref{eq:xy_Langevin_int})
in time via the Euler-Maruyuma method \cite{maru}.  The resulting algorithm is essentially an application of
``Brownian dynamics with hydrodynamic interactions" \cite{mccammon} applied
to membrane shape fluctuations and protein motion.  Adopting the 
envelope function defined in Eq. (\ref{eq:delta})  allows a seamless melding of the approaches introduced in Refs.  \cite{Naji,Seifert_pre,Brown_prl} 
(a detailed description of our algorithm will be provided in a future publication). 
%(a detailed description of our algorithm will be provided elsewhere \cite{in_prep}).

%In the simulations we use $L=250$~nm and $M=100$ giving $a=2.5$~nm which is comparable to 
%bilayer thickness. The protein area is chosen as $A_{\mathrm{p}}=(4a)^2=100\,{\mathrm{nm}}^2$
%giving a size (diameter) of $a_{\mathrm{p}}\sim10$~nm.
%Membrane bending rigidity is set to $K_{\mathrm{m}}/(k_{\mathrm{B}}T)=5$, temperature to $T=300{\mathrm K}$ 
%and the bare diffusion coefficient for the protein to $D_0 = 5.0$~$\mu{\mathrm m}^2/{\mathrm s}$, qualitatively reflecting the motion of band 3 protein dimers on the surface of human red blood cells (RBC) \cite{Brown_prl}.  
%We allowed the viscosity
%of the medium, the protein bending modulus  and the protein spontaneous curvature to
%vary over the ranges  $\eta=0.001$ (water) to 0.006 ${\mathrm{Pa}}\cdot{\mathrm{s}}$ (RBC cytoplasm),
%$K_{\mathrm{p}}/K_{\mathrm{m}}=1$ to 8 and  $C_{\mathrm{p}} = 0.025\,{\mathrm{nm}}^{-1}$ to $0.15\,{\mathrm{nm}}^{-1}$ respectively.  
%For simplicity it was assumed that both bilayer and protein
%saddle-splay moduli were equal in magnitude but opposite in sign to the corresponding curvature moduli, i.e. $K'_{\mathrm{m(p)}}=-K_{\mathrm{m(p)}}$.
%The projected diffusion coefficient, $D$, is calculated as  
%$D = \lim_{t\rightarrow\infty} \overline{[{\mathbf r}(t)-{\mathbf r}(0)]^2}/(4t)$. The simulations run for
%$\sim5\times 10^7$ time steps of size $\sim0.01$~ns, ensuring both numerical
%accuracy and statistically reliable results.

%%%------Fig surface projection 
\begin{figure}[t!]
\begin{center}
\vspace{-.6cm}
\includegraphics[angle=0,width=7.2cm]{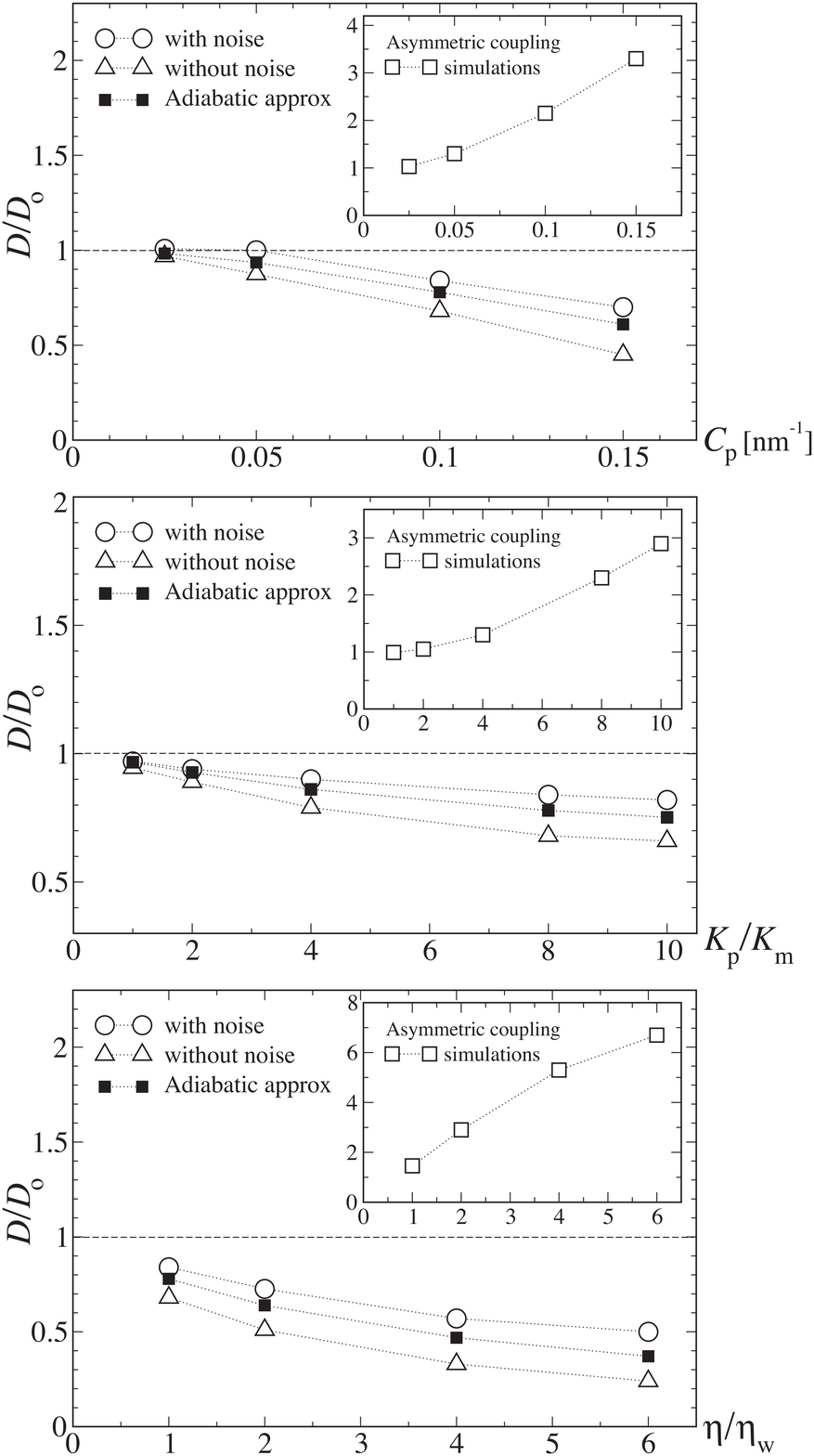}
\caption{Projected diffusion coefficients for protein motion on an elastic membrane.  
%``Default" parameters for all panels include the protein elastic properties 
 %$K_{\mathrm{p}}=-K'_{\mathrm{p}}=40\,k_{\mathrm{B}}T$
%and $C_{\mathrm{p}} =0.1\,{\mathrm{nm}}^{-1}$ and viscosity of the surrounding
%solvent $\eta = \eta_{\mathrm{w}}=0.001~{\mathrm{Pa}}\cdot{\mathrm{s}}$ (water).  In each panel,
%one of these three parameters is varied as indicated while holding the other two at the default values
%(see text for other parameters).
In each panel, one of the three parameters $K_{\mathrm{p}}$ (protein bending modulus), $C_{\mathrm{p}}$ (protein spontaneous curvature),
and $\eta$ (medium viscosity) is varied as indicated while holding the remaining physical properties at the default values
indicated in Table \ref{tab:sim_para}.
 Circles indicate the results of full simulations including thermal motion of both the bilayer and the protein.  
 The triangles indicate results obtained by turning off all bilayer shape
 fluctuations (i.e. setting $T=0$ for membrane undulations modes) 
 and the solid squares indicate the results of the adiabatic theory discussed in the text.  
Insets show simulation results obtained within the asymmetric coupling approximation. 
%N=500
Errorbars are approximately the size of the symbols. 
Diffusion coefficients are calculated as  
$D =  \overline{[{\mathbf r}(t)-{\mathbf r}(0)]^2}/(4t)$. The simulations run for
$\sim 5\times 10^7$ time steps of size $\Delta t\sim 0.01$ns, ensuring both numerical
accuracy and statistically reliable results.  The protein's root-mean-square displacements over the course of the
simulations are approximately 100nm (i.e. 20 times the protein's radius). 
}
\vspace{-1cm}
\label{fig:Fig_varCp}
\end{center}
\end{figure}

The protein's influence on the membrane is most easily seen in the absence of thermal shape fluctuations of the membrane, but for sufficiently large $C_{\mathrm{p}}$ and $K_{\mathrm{p}}$ values the effect of the protein remains clearly visible despite thermal fluctuations of the membrane
surface (see Fig. \ref{fig:snapshot}).
The average distortion of the membrane surrounding the protein is sufficient to significantly slow
protein motion in all cases we have studied (see Fig. \ref{fig:Fig_varCp}).  This slowing derives
from two effects.  First, the protein tends to trap itself in the deformation created by its own perturbation
to the bilayer.
The energy-minimized configuration displayed in Fig. \ref{fig:snapshot} places the protein at the
bottom of a curved valley; attempted diffusion up the walls of this valley is hindered by the interaction-induced force in Eq. (\ref{eq:xy_Langevin_int}).  
Second, $x,y$ translation of the protein is accompanied by  translation
of the membrane deformation surrounding this protein.  The hydrodynamic effects included in Eq.
(\ref{eq:q_memb_langevin}) dictate that the translation of such a deformation is resisted by
the viscous drag of the medium surrounding the bilayer.  This drag acts in addition to the 
usual quasi-2D drag incorporated within $D_0$ and slows protein motion.  
These effects are most pronounced when shape fluctuations of the bilayer are neglected 
by setting $T=0$ in Eq. (\ref{eq:q_memb_langevin}).  Increasing the local
deformation around the protein or the solvent viscosity both decrease
$D$.  In Fig. \ref{fig:Fig_varCp} we display three means to control
this effect.  Increasing $C_{\mathrm p}$ drives large deformations from a
flat plane for any finite $K_{\mathrm p}$.  Larger values of $K_{\mathrm p}$ will tend to increase
this effect up until the point where the rigidity of the protein becomes effectively infinite and
the response to the protein saturates.  The hydrodynamic drag in our model is
controlled via $\eta$; increases in $\eta$ reduce $D$ as effectively
as do perturbations to membrane shape.

Within the asymmetric coupling approximation, which ignores the influence of
the protein on membrane shape, it is predicted that bilayer shape fluctuations will enhance 
curved protein mobility \cite{Seifert,Seifert_crv} (i.e. $D>D_0$, insets Fig. \ref{fig:Fig_varCp}).  Although this effect can be seen in our simulations, the enhancement is slight (relative to the similar effect
within the aforementioned approximation scheme) and can not overcome the dominant slowing
caused by the protein's distortion of the bilayer (compare open circles and triangles, Fig. \ref{fig:Fig_varCp}).  We find $D<D_0$ for all cases studied, which represents a qualitative departure from earlier predictions.

%Although thermal shape fluctuations
%of the bilayer do raise $D$ above the idealized results neglecting these fluctuations, the effect
%is quite small and it is always the case that $D<D_0$.
%(The purely geometric effect due to projection of protein motion to the base plane \cite{Halle_NMR,Seifert,Seifert_pre,Gov_pre,Naji} is negligible for the cases considered
%here.)

We may approximately account for the viscous drag effect discussed above by
invoking an adiabatic approximation and assuming that the energy-minimized
membrane distortion profile (denoted by $\bar h({\mathbf x}, t)$)
instantaneously tracks protein position. % \cite{note}.
%We denote this instantaneously minimized shape profile as $\bar h({\mathbf x}, t)$.  
Hence, 
\begin{equation}
  \dot {\bar h}({\mathbf x}, t) = -{\mathbf v}\cdot\nabla \bar h({\mathbf x}, t), 
  \label{eq:translation}
\end{equation}
for a protein moving at constant lateral velocity ${\mathbf v}$. The power dissipated by 
viscous losses in the medium may be calculated from 
 $ P = %\int_{{\mathcal A}_\bot}\! {\mathrm{d}}{\mathbf x}\, \dot {\bar h}({\mathbf x}, t) \, F({\mathbf x}, t) 
     \frac{1}{L^2}\sum_{{\mathbf q}} \dot {\bar h}_{{\mathbf q}}(t) F_{-{\mathbf q}}(t)$,
 where $ F_{{\mathbf q}}(t)$ follows from Eq. (\ref{eq:q_memb_langevin}) as
 $ F_{{\mathbf q}}=\dot {\bar h}_{{\mathbf q}}/ \Lambda_{{\mathbf q}}$. Thus,
 by using Eq. (\ref{eq:translation}), 
 the power loss may be written as
   $P =  %\frac{1}{L^2}\sum_{{\mathbf q}} \frac{(\dot {\bar h}_{{\mathbf q}})^2}{ \Lambda_{{\mathbf q}}}
        \frac{1}{L^2}\sum_{{\mathbf q}}  ({\mathbf v}\cdot{\mathbf q})^2|\bar h_{{\mathbf q}}|^2/ \Lambda_{{\mathbf q}}\equiv|{\mathbf v}|^2/\mu_{\mathrm{def}}$ with $\mu_{\mathrm{def}}$ being
the effective mobility of the deformation. The effective protein diffusion coefficient follows 
from $D_{\mathrm{def}} = \mu_{\mathrm{def}} k_{\mathrm{B}}T$ and $D^{-1} = (D_0^{-1} + D_{\mathrm{def}}^{-1})$ to give
\begin{equation}
\frac{1}{D} \approx \frac{1}{D_0}+\frac{2\eta}{k_{\mathrm{B}}TL^2}\sum_{{\mathbf q}} |{{\mathbf q}} |^3 |\bar h_{{\mathbf q}}|^2.
\end{equation}
This expression depends only on the minimized deformation profile of the 
membrane at fixed protein position and is readily calculated (shown as solid squares in Fig.  \ref{fig:Fig_varCp}).
Although the approximation is imperfect due to neglect of membrane %shape 
fluctuations and the self-trapping effect discussed above, the adiabatic results serve as a reliable estimator of the observed 
trends for the full simulation.

We are not aware of experimental studies that specifically investigate
the role of protein curvature on self-diffusion, but
do note that recent experiments \cite{urbach} show deviations
from the standard theory \cite{Saffmann} used to predict 
membrane-protein mobility.  The effect described here may prove to be 
important in describing these deviations for certain proteins.  The
solvent viscosity dependence we find is at odds with the 
weak (logarithmic) dependence expected for flat proteins \cite{Saffmann} 
and provides a concrete means to verify our predictions experimentally.
%We have applied our approach to a single application in this preliminary
%study, but emphasize that the numerical techniques developed in this work are
%quite general.  Identical simulation methodologies may be used to explore interactions
%between membranes and cytoskeletal filaments, active proteins \cite{Girard}, polymers
%and other sources of dynamic influences on the membrane surface.  

This work is supported by the NSF (CHE-034916, CHE-0848809, DMS-0635535), the BSF 
(2006285) and the Camille and Henry Dreyfus Foundation.  We thank N. Gov, H. Diamant and H. Boroudjerdi
for helpful discussions.

\vspace{-.5cm}
%%%%%%%%%%%%%%%%%%%%%%%%%%%%%%%%%%%%%%%%%%%%%%%%%%%%%%%%%%%%%%%%%%%%%%%%%%

\end{document}